%% file: icc2019.tex
\documentclass[conference,a4paper,final]{IEEEtran}
\IEEEoverridecommandlockouts

\input{imports}
\input{commands}

\usepackage[utf8]{inputenc}

\def\BibTeX{{\rm B\kern-.05em{\sc i\kern-.025em b}\kern-.08em
    T\kern-.1667em\lower.7ex\hbox{E}\kern-.125emX}}

\newlength\figureheight
\newlength\figurewidth
\pgfplotsset{every axis/.append style={
                legend style={font=\footnotesize},
                ylabel near ticks,
                xlabel near ticks,
                }
            }

\DeclareSIUnit\dBW{dBW}
\DeclareSIUnit\dBi{dBi}
\DeclareSIUnit[per-mode=symbol]{\bps}{\bit\per\second}
\DeclareSIUnit[per-mode=symbol]{\Gbps}{\giga\bit\per\second}

\input{gloss}

\begin{document}

\title{Multi-Satellite Multi-User MIMO Precoding: Testbed and Field Trial \\
\thanks{This work has been supported by the German Aerospace Center (DLR), Space Administration, and the German Federal Ministry for Economic Affairs and Energy (BMWi) under grant number 50 YB 1618. We express our special thanks to Eutelsat S.A.\ for providing us with satellite capacity (E7A/E7B).}
}

\author{\IEEEauthorblockN{Kai-Uwe~Storek, Robert~T.~Schwarz, and~Andreas~Knopp}
\IEEEauthorblockA{Chair of Signal Processing, Bundeswehr University Munich, 85579 Neubiberg, Germany\\
Email: papers.sp@unibw.de}
}

\maketitle

\begin{abstract}
Precoding for multibeam satellite systems with \acrlong{FFR} in a \gls{MU-MIMO} downlink scenario is addressed. A testbed is developed to perform an over-the-air field trial of \acrlong{ZF} precoding for the spatial multiplexing of two separate video streams over two co-located geostationary satellites. Commercial-off-the-shelf DVB-S2x receivers in two single-antenna \glspl{UT} are operated to successfully decode two independent video streams. To this end, particular attention is paid to a comprehensive assessment of the practical synchronization tasks of a precoding based \gls{MU-MIMO} system. In particular, carrier frequency recovery and carrier phase synchronization in the most challenging multi-satellite scenario with different oscillators in the payloads is performed. An estimation method for \gls{CSI}, i.e. the complex channel coefficients, is also proposed and implemented. The successful video transmission finally constitutes the first field trial of MU-MIMO precoding and proves the feasibility of precoding concepts for multibeam satellite systems. 
\end{abstract}

\glsresetall


\section{Introduction}

Precoding for multibeam satellite systems is a promising technique to further push the throughput of \gls{VHTS} systems. 
In \gls{SATCOM}, the precoding of information at the gateway is usually preferred over equalization strategies at the receiver in order to keep the complexity of the \glspl{UT} or the satellite payload low. Particularly in a \gls{FFR} scenario, the interference between neighboring beams can effectively be suppressed by precoding of the transmit signals such that the \gls{SINR} for the individual users is increased. Transmit signal preprocessing requires \gls{CSI} at the transmitter, and the common assumption in \gls{SATCOM} is to perform a joint processing in the gateway. This allows single antenna \glspl{UT} without the need of cooperation and enables simple receiver hardware.

Based on the basic theory of non-linear and linear precoding techniques \cite{Joham2005, Weingarten2006}, several approaches have been developed for multibeam satellite systems. Their objectives differ according to the satellite payload architecture which can either use a single or multiple reflectors to generate the beams.
Due to the very long satellite-to-ground-distance compared to the feed separation in one feed cluster, the feeds illuminating a single reflector appear as a single point in the far field. As a consequence, single reflector designs cannot take advantage of conventional spatial diversity schemes \cite{Perez-Neira2019}.

The precoding can only exploit the fact that, due to the gain patterns, users will receive signals with different amplitudes from the feeds,
and the basic objective of precoding is interference mitigation or interference alignment. A prominent approach, which is also known from terrestrial cellular networks, is the multicast precoding \cite{Christopoulos2015}. It basically aims to form groups of users that are close together, which is also known as geoclustering \cite{Guidotti2017}, so that the mutual interference of the group members cannot be distinguished at the satellite due to their very similar or even identical channel vectors. Information for the group members is then embedded in the same frame while interference mitigation is performed for different geoclusters.

On the other hand, the multiple-reflector approach has the potential to additionally exploit the signal phase in the channel vectors. This provides a further degree of freedom to apply \gls{SDMA} with spatial multiplexing and unicast transmission. A comprehensive overview of \gls{MIMO} techniques with spatial multiplexing can be found in \cite{Schwarz2019}, where it has also been reported that multiple-reflector schemes outperform the single-reflector designs. The spatial distribution of the antenna reflectors can be performed in a single-satellite scenario or in a multiple-satellites scenario. In the single-satellite scenario, all reflectors are on a single spacecraft like, for example, the four side deployable reflectors of Eutelsat's KA-SAT \cite{Fenech2016}. In the more challenging multiple-satellites scenario, the reflectors are on different spacecrafts which themselves are either in the same orbital slot (so-called satellite co-location \cite{Soop1994}) or even at very different orbit positions. The latter case is particularly suited for lower frequency bands such as UHF, where substantial multiplexing gains have already been reported and demonstrated in practice \cite{Hofmann2017}. The co-located satellites scenario has been theoretically evaluated in \cite{Schwarz2019a} but not yet practically demonstrated. However, from a commercial perspective, the operation of co-located spacecrafts in a \gls{FFR} scheme is of great interest because it enables the concurrent operation of smaller spacecrafts in more cost-efficient redundancy concepts.

At the same time, among all proposed applications, the practical realization of precoding with multiple satellites is the most challenging task due to manifold synchronization and estimation problems. For example, precoding in the downlink requires a phase controlled radiation of the multiple signal streams to the users. An important task is, therefore, the synchronization of the multiple free-running oscillators on different payloads. Potential \glspl{CFO} must be tracked and compensated at the transmitting gateway. Moreover, multiple satellites move independently in their orbits and show slight relative motions, which results in varying Doppler shifts as well as varying channel states. The estimation of the \gls{MU-MIMO} channel, the compensation of the mentioned effects as well as the update rate of the \gls{CSI} at the gateway depending on the application scenario are of utmost importance. Here, the CSI estimation problem comprises the channel amplitude and phase.

While precoding based \gls{MU-MIMO} transmission has been thoroughly discussed under perfect theoretical assumptions with respect to the \gls{CSI} and synchronization accuracy, practical testbeds or field trials are still lacking. However, it is well known that the performance of precoding is highly sensitive to inaccuracies in the CSI estimation as well as to synchronization errors. In the satellite-to-ground channel, the update rate of the CSI estimates and the synchronization loop bandwidth are strongly limited by the \gls{RTT} of the signals. Finally, when considering the precoding of both the signal amplitude and phase while using spatially separated antennas, the true spatial correlation in the channel and the rate of change in the atmospheric impairments are unknown error sources.

In this context, the contribution of this paper is to show and discuss the results of the first practical testbed and field trial of a precoding based \gls{MIMO} signaling scheme employing two co-located satellites. Our objective is to use commercially available DVB-S2x equipment at the \glspl{UT} and put all the necessary layers of signal processing on top to enable the transmission of two independent video streams in an \gls{FFR} \gls{MIMO} mode. The rationale behind this procedure is basically that the commercial equipment will only be able to decode the streams if the \gls{MIMO} provides the needed channel capacity, the above mentioned estimation and synchronization tasks have been thoroughly performed, and the precoding has been implemented correctly to cancel out any interference. A thorough theory of the synchronization accompanies our contribution.

The paper is structured as follows: After introducing the demonstration setup with two co-located satellites and the system model of the \gls{MU-MIMO} downlink scenario in Section\ \ref{sec:TestbedAndSystemModel}, we theoretically discuss and present solutions for the practical aspects of the involved synchronization and estimation problems in Section\ \ref{sec:Challenges}. The 
results of the field trial are then presented and discussed in Section\ \ref{sec:Results}.

\section{Testbed and System Characterization}
\label{sec:TestbedAndSystemModel}

\subsection{Trial Overview}
To demonstrate precoding over multibeam satellites, we selected the co-located geostationary satellites "EUTELSAT 7A" (E7A) and "EUTELSAT 7B" (E7B) at 7\unit{\degree}E. Both satellites provide coverage in Central Europe. Since some transponders on both satellite cover the same frequency band and polarization in the downlink, only one of them can usually be used. A simultaneous and uncoordinated usage of these transponders lead to strong interference on ground prohibiting an operational service.
In contrast, if the satellites are used in a coordinated way with signal preprocessing at the gateway station, a significant throughput gain is expected in theory \cite{Schwarz2019}. When multiple co-located satellites shall be collaboratively used for operational services, special requirements with respect to the co-location strategy and station-keeping methods have to be considered. A comprehensive analysis about these requirements is provided in \cite{Schwarz2019a}.
In our demonstration campaign, we used both satellites to set up a \gls{MU-MIMO} \gls{SATCOM} system. The necessary signal preprocessing has been performed at the gateway station. Due to slightly different conversion frequencies of the payloads, the signals in the uplink are separated in frequency, but the downlink uses the same part of the frequency band and, hence, \gls{FFR} is realized. The scenario combines a \gls{SISO} uplink with a \gls{MIMO} downlink (see Fig.\ \ref{fig:scenario}), which is the common assumption for recent HTS satellite systems \cite{Perez-Neira2019,Joroughi2017}.
Three antennas have been involved in our trial (see Fig.\ \ref{fig:antennafarm}). Both uplink carriers were anchored by a 4.6\unit{\m} gateway antenna. Dishes of diameter  7.6\unit{\m} at (48.073556\unit{\degree}N, 11.63054633\unit{\degree}E) and 1.8\unit{\m} at (48.073395\unit{\degree}N, 11.6308\unit{\degree}E) acted as receiving antennas for the \glspl{UT}. We uplinked two DVB-S2 modulated carriers with symbol rate of 1.25\unit{\mega baud} and roll-off equal to 0.2 for video transmission. The modulation and coding scheme was QPSK 5/6.
%
%
\begin{figure}[t]
\centerline{
	\input{pics/scenario}
	}
\caption{Demonstration setup showing the gateway station, two co-located satellites and two UTs. Uplink signals are separated in frequency (two separate links), downlink uses same frequency and polarization (\gls{MU-MIMO} channel).}
\label{fig:scenario}
\end{figure}
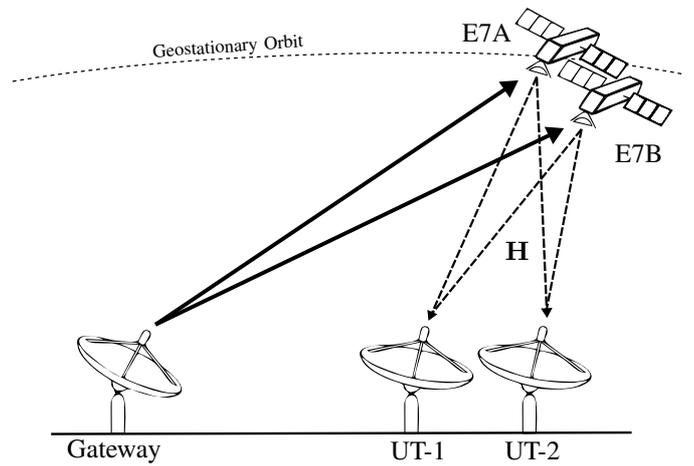
\begin{figure}[t]
\centerline{
	\input{pics/antennafarm.tex}
	}
\caption{Picture showing the involved ground antennas located at the Munich Center for Space Communications in the city of Munich.}
\label{fig:antennafarm}
\end{figure}
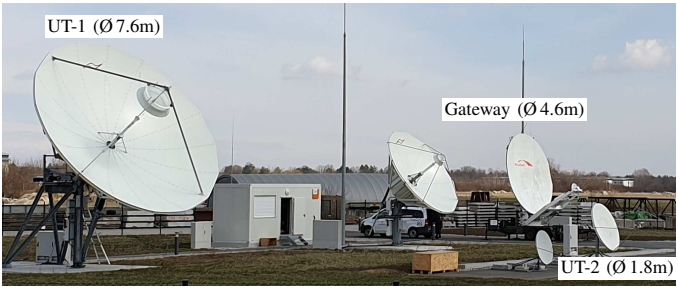
\subsection{Transmitter}
\begin{figure}[tb]
\centering
\input{pics/sdr_impl}
\caption{Block diagram of the relevant parts of transmit SDR implementation. Two MPEG transport streams are modulated according to the DVB-S2(x) standard. ZF precoding is applied by matrix $\precodM$, which is calculated based on \gls{CSI} provided by the \glspl{UT}. CFO compensation is done by estimation and inversion of $\TCFO$. One reference tone for each satellite is generated and received at the gateway (closed-loop) to estimate the entries of $\TCFO$.}
\label{fig:sdr_impl}
\end{figure}
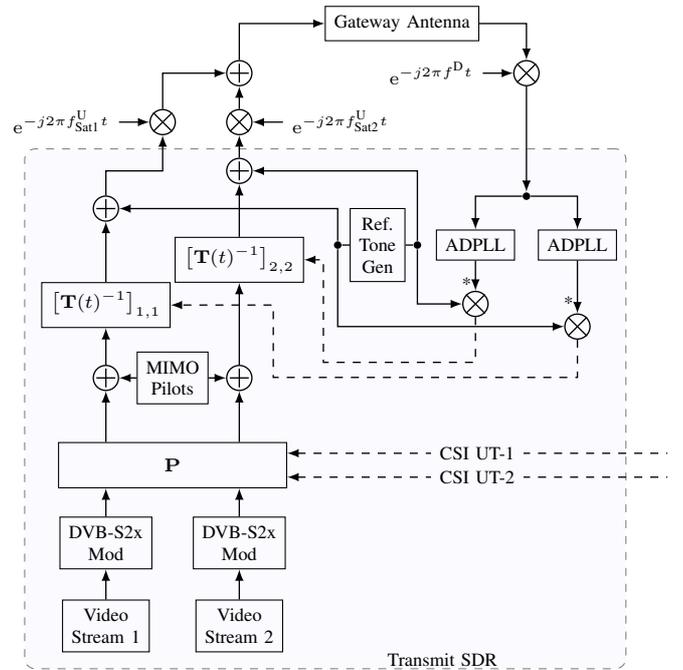

A \gls{SDR} has been used in the gateway for transmit signal preprocessing. A block diagram showing the main parts of the implementation is given in \mbox{Fig.\ \ref{fig:sdr_impl}}. After video encoding and framing, the bit streams are modulated according to the DVB-S2 standard. The symbol streams are then input to the precoding stage. The calculation of the precoding matrix $\precodM$ will be discussed in Section\ \ref{sec:sysmodel}.  
To determine $\precodM$, the individual channel vectors are estimated at the \glspl{UT} and fed back using a terrestrial link.
To enable the estimation of \gls{CSI} at the \glspl{UT}, non-precoded \gls{MIMO}  pilots are added. Comparable to terrestrial mobile \gls{MIMO} communication, we set up a separate channel for these pilots. Since all \glspl{UT} can acquire \gls{CSI} by this single channel, this technique is termed as \gls{CRS} \cite{Clerckx2013} and limits the overhead. Orthogonal Zadoff–Chu sequences consisting of 2000 symbols each and occupying a bandwidth of 200\,kHz were used as \gls{CRS} during the trial.\footnote{Optimization of \gls{CRS} parameters with respect to the bandwidth efficiency were not subject of this trial. For future \gls{MU-MIMO} \gls{SATCOM} applications, the parametrization and arrangement of the \gls{CRS} has to be adopted and optimized from a system perspective.}
After the precoding matrix $\precodM$ is applied and the pilots are added, the precompensation of the oscillator offsets is performed. We used an \gls{ADPLL} at the gateway station to track reference tones that were transmitted via both satellites. The difference between the transmitted and the received reference tones is used as $\TCFO^{-1}$ to precompensate the oscillator offsets in a computationally efficient manner. A detailed description of this \gls{CFO} compensation technique is presented in Section\ \ref{sec:Challenges}.

The reference tones themselves as well as the MIMO pilots are added spectrally separated from the precoded carriers (``out-of-band'') before the two baseband signals are shifted to their individual uplink frequencies at 13\,GHz by a common upconverter.
\subsection{Receiver}
\begin{figure*}[tb]
\centering
\input{osci_chain}
\caption{Overview about the testbed and the sources for frequency shifts.}
\label{fig:osci_chain}
\end{figure*}
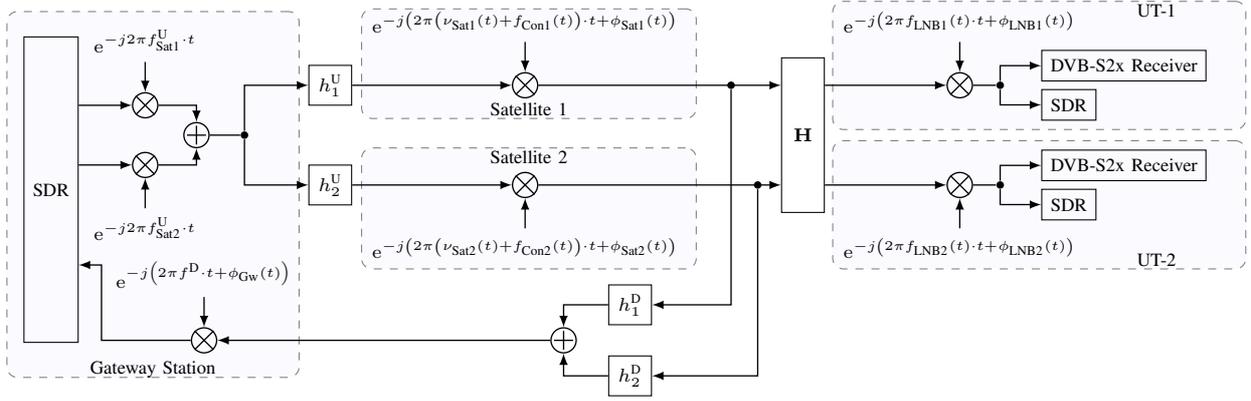

The two uplink signals are mixed to the same downlink frequency $\DownlinkFreqOne$ at 11.5\,GHz.  Each \gls{UT} converted the received signal to L-band intermediate frequency, where the signal has been split (see also Fig.\ \ref{fig:osci_chain}). Conventional DVB-S2x demodulators (Ericsson RX8200) are used as the main receiving unit in each \gls{UT}. These demodulators demonstrate the compatibility of \gls{COTS} components with MU-MIMO SATCOM systems since no modification has been made to them. Furthermore, one \gls{SDR} is installed in parallel to each DVB-S2x demodulator. The task of the receive \glspl{SDR} is the estimation of the \gls{CSI} and the transfer of these channel vectors back to the transmit SDR since this kind of channel sounding is not part of DVB-S2x standard. We employed a best linear unbiased estimator comparable to the estimator used in \cite{Hofmann2016}. The update rate of the CSI was set to \nicefrac{1}{5}\,Hz, whereas five \gls{CSI} measurements were executed and averaged before a message with the \gls{CSI} was sent to the transmit SDR.
%

\subsection{System Characterization}
\label{sec:sysmodel}
We briefly recapitulate a simplified system characterization to point out the assumptions about the propagation channel and the realization of precoding that is executed at the gateway station. The focus is on the important downlink part that represents the MIMO SATCOM channel. This simplification is reasonable since the uplink is realized in \gls{SISO} mode and the link budget has been downlink limited with respect to the achievable \gls{CNR}. In Sec.\ \ref{sec:Challenges} the model will be extended to cover additional frequency offset effects.

Considering a non-cooperative \gls{UT} with a single antenna, the link between the $\Ntran$ satellites and one \gls{UT} on Earth represents a \gls{MISO} channel. The received signal of the $k$-th \gls{UT} is given by
\begin{align}
	\recSymbol_{k} = \Transpose{\channelVec}_k \sendVec + \noiseSymbol_k
	\label{eqn:receive_symbols_single_user}
\end{align}
where $\channelVec_k \in \complexSet{\Ntran \times 1}$ represents the channel coefficients between the $\Ntran$ satellite antennas and the $k$-th receive antenna. The variable $\sendVec \in \complexSet{\Ntran \times 1}$ is the vector that comprises the transmit symbols and $\noiseSymbol_k \in \complexSet{\Ntran \times 1}$ is complex circular symmetric noise with zero mean and variance $\sigma^2_k$.

The channel coefficients in $\channelVec_k$ are defined by a frequency flat channel model with  \gls{LOS} propagation characteristics
\begin{align}
	\channelSymbol_{k,n} = \frac{\speedOfLight}{4 \pi \DownlinkFreqOne \antennaDist{k}{n}} \cdot \Euler^ { -\complexLetter \frac{2 \pi \DownlinkFreqOne}{\speedOfLight} \antennaDist{k}{n} } .
	\label{eqn:calcOfCtmt}
\end{align}
The parameter $\DownlinkFreqOne$ is the downlink carrier frequency and  $\speedOfLight$ the speed of light. The variable $\antennaDist{k}{n}$ is the distance between the $k$-th \gls{UT} and the $n$-th satellite.

Serving $\Nusers$ single-antenna receivers at the same time defines the \gls{MU-MIMO} system, where $\Nusers=2$ in our trial. Together with the $\Ntran=2$ satellites, the \gls{MU-MIMO} system can be described by
\begin{align}
	\recVec & = \ctm \sendVec + \noiseVec           \\
	        & = \ctm \precodM \dataVec + \noiseVec,
	\label{eqn:mimoSystemModell}
\end{align}
where $\dataVec = \Transpose{(\dataSymbol_1, ..., \dataSymbol_K)} \in \complexSet{\Nusers \times 1}$ is a vector containing the $\Nusers$ data symbols. These symbols are i.i.d. zero mean complex Gaussian random variables with unit variance. The variable $\precodM = \left(\precodVec_1,...,\precodVec_K \right) \in \complexSet{\Ntran \times \Nusers}$ denotes the precoding matrix. The channel matrix is defined by $\ctm = \Transpose{\left( \channelVec_1, ..., \channelVec_K\right)} \in \complexSet{\Nusers \times \Ntran}$, the noise vector by $\noiseVec = \Transpose{\left(\noiseSymbol_1, ..., \noiseSymbol_K \right)} \in \complexSet{\Nusers \times 1}$  and the vector of the received symbols by $\recVec = \Transpose{\left(\recSymbol_1,...,\recSymbol_K\right)} \in \complexSet{\Nusers \times 1}$.
\newcommand{\ZFScaleMat}{\Matrix{\Lambda}}
The task of the precoding matrix $\precodM$ is to control the multiuser interference. A common linear approach to calculate $\precodM$ is \gls{ZF}. With \gls{ZF} the interference is completely canceled and the \glspl{UT} see only their intended signals. Assuming $\ctm$ is of rank $\Nusers$, $\precodM$ is calculated such that $\ctm \precodM = \ZFScaleMat^{1/2} = \diag{\sqrt{\lambda_1},\dots,\sqrt{\lambda_k}}$ with
\begin{align}
	\precodM = \Hermitian{\ctm}\left( \ctm \Hermitian{\ctm} \right)^{-1} \ZFScaleMat^{1/2}.
	\label{eqn:calcOfP}
\end{align}
Matrix $\ZFScaleMat$ is chosen such that
\begin{align}
	\MatrixElement{\precodM \Hermitian{\precodM}}{n}{n} \leq 1,
\end{align}
\newcommand{\sumRateZF}{R_\text{sum}^{\text{ZF}}}
in order to ensure a \gls{APC}. After \gls{ZF} precoding, the receiving signal at UT $k$ becomes $\recSymbol_{k} = \sqrt{\lambda_k} \dataSymbol_k + \noiseSymbol_k$, and the sum rate of the system is given by
\begin{align}
	\sumRateZF = \sum^{K}_{k=1}\left( \log_2\left( 1 + \frac{\lambda_k}{\sigma^2_k}\right) \right).
	\label{eqn:sumrate}
\end{align}
\section{Frequency Synchronization in Distributed MIMO SATCOM Systems}
\label{sec:Challenges}
The coordinated usage of geographically displaced transmit antennas is termed as a distributed MIMO system. An essential requirement for effective precoding is carrier phase synchronization. For terrestrial distributed MIMO systems, different realizations exist in order to synchronize all involved \glspl{AP} \cite{Balan2013,Rahul2012}. To the best of our knowledge, all the available approaches rely on synchronization procedures, where pilots, transmitted from a master \gls{AP}, are used by the slave \glspl{AP} to adjust their individual frequency and phase deviation. This approach cannot be transferred to a multi-satellite precoding system, since no inter-satellite links are available. Furthermore, most satellites in orbit do not provide programmable digital payloads for signal processing tasks. Before going into the detail how synchronization is achieved, we first want to point out the sources for \glspl{CFO} which are visualized in Fig.\ \ref{fig:osci_chain}.

\gls{CFO} contributions are not expected at the gateway station since all components can be easily synchronized by reference clocks. However, phase noise is introduced by the common Ku-band up- and also by the downconverter. Both effects are summarized in $\phi_\text{Gw}(t)$ for our scenario. Regarding the actual data transmission, a phase shift introduced by the upconverter at the gateway station has no impact on the precoding performance as it is a common phase shift equally effecting all precoded symbols.
In contrast, the signal relaying by the space segment causes multiple frequency shifts. Both satellites are permanently moving along their trajectories given by the co-location method.  From a fixed perspective on Earth, the actual distance to a satellite in \gls{GEO} alternates within a 24h time frame. Based on the relative speed $v(t)$ of a satellite with respect to the transmit or receive antennas on Earth, the time-variant relativistic Doppler shift $\dopplerFreq(t)$ can be calculated  with
\begin{align}
	\dopplerFreq(t) = f \sqrt{\frac{c+v(t)}{c-v(t)}}.
	\label{eqn:relaDoppler}
\end{align}
Using (\ref{eqn:relaDoppler}) to assess the \gls{CFO} contribution due to the Doppler effect for Satellite 1, yields
\begin{align}
	\DopplerOne = \left( \UplinkFreqOne + \DownlinkFreqOne \right) \cdot \sqrt{\frac{c+v_{\text{Sat1}}(t)}{c-v_{\text{Sat1}}(t)}},
\end{align}
where $\UplinkFreqOne$ denotes the uplink carrier frequency for the satellite 1 and $\DownlinkFreqOne$ is the common carrier frequency of the downlink.  We simulated $\DopplerOne$ and $\DopplerTwo$ by the help of public available ephemeris data. The sinusoidal curves of $\DopplerOne$ and $\DopplerTwo$ have a periodicity of about 24h with amplitudes in the order of $\pm$150\,\unit{\hertz} with respect to antenna locations in Central Europe.

Another source for frequency shifts at the space segment are the frequency converters used to shift the \gls{RF} signal from the uplink into the downlink band. These converters are not synchronized to an external reference, and frequency drifts, for example due to aging effects, are the consequence. A typical requirement regarding the stability of a satellite frequency converter is in the order of 1\,ppm per year life-time. Since Ku-band payload converters have nominal frequencies around 2\unit{\giga\hertz}, we expect additional \gls{CFO} contributions by the converters, defined as $\ConvSatOne$ and $\ConvSatTwo$, to be lower than 20\unit{\kilo\Hz}. In addition, every converter also adds phase noise, $\pnSatO(t)$ and $\pnSatT(t)$ to the signal.

Finally, the \glspl{LNB} of the \glspl{UT} are another source for \gls{CFO} and are defined by the variables $\LnbFreqUserOne$ and $\LnbFreqUserTwo$. The stability of the oscillators integrated in the user equipment differ substantially, and offset values in Ku-band can exceed 1\unit{\mega\Hz}.
The time-invariant channel model in (\ref{eqn:calcOfCtmt}) is now extended by the time-variant oscillator offsets resulting in
\begin{align}
	\ctmt(t) = \RCFO \ctm \TCFO
\end{align}
with
\begin{align}
\TCFO = \text{diag}  \Big(  &\Euler^{ -\complexLetter (2 \pi \left(\DopplerOne + \ConvSatOne\right)\cdot t + \pnSatO(t))},  \nonumber\\
			  		  		&\Euler^{ -\complexLetter (2 \pi \left(\DopplerTwo + \ConvSatTwo\right)\cdot t + \pnSatT(t))} 	\Big)
\end{align}
and
\begin{align}
	\RCFO = \text{diag}  \Big(  	&\Euler^{ -\complexLetter( 2 \pi \LnbFreqUserOne \cdot t + \phi_\text{LNB1}(t))},  \nonumber\\
			  		  				&\Euler^{ -\complexLetter( 2 \pi \LnbFreqUserTwo \cdot t + \phi_\text{LNB2}(t))} 	\Big)
\end{align}
It is becomes apparent that without \gls{CFO} compensation a symbol transmission via
\begin{align}
		&\ctmt \precodM = \RCFO \ctm \TCFO \precodM
\end{align}
is not longer diagonal since $\precodM$ is based on $\ctm$. Diagonalization can be achieved by precompensation of $\TCFO$ through its inverse at the the gateway station, which leads to
\begin{align}
	\recVec &= \ctmt \TCFO^{-1} \precodM \dataVec + \noiseVec 				\nonumber \\
	        &= \RCFO \ctm \TCFO \TCFO^{-1} \precodM \dataVec + \noiseVec 	\nonumber \\
	        &= \RCFO
	        \begin{bmatrix}
				\sqrt{\lambda_1} & 0 \\
				0 & \sqrt{\lambda_1}
				\end{bmatrix} \dataVec + \noiseVec
\label{eqn:cfoPrecod}
\end{align}
Beside the fact, that the CFO precompensation is possible at the gateway station, (\ref{eqn:cfoPrecod}) also shows that frequency offsets at the receivers do not influence the sum rate because of its diagonal structure, $\RCFO$ only adds a frequency offset to the already superpositioned signals. For
\begin{align}
	\TCFO^{-1} = \text{diag}  \Big( &\Euler^{ \complexLetter ( 2 \pi \left(\DopplerOne + \ConvSatOne\right)\cdot t +   \pnSatO(t))}, \nonumber\\
									&\Euler^{ \complexLetter (2 \pi \left(\DopplerTwo + \ConvSatTwo\right)\cdot t + \pnSatT(t))} \Big)
\end{align}
estimates for $\DopplerOne + \ConvSatOne$ and $\DopplerTwo + \ConvSatTwo$ are required. The compensation of the phase noise is nearly impossible as a consequence of the long \gls{RTT} of 250\,ms. Due to this long propagation time from the origin of the phase noise until the arrival of the precompensated signal, only offset frequencies below \nicefrac{1}{4}\,Hz are compensable. Such low offset frequencies are indistinguishable from slow varying frequency drifts and are therefore modeled in $\ConvSatOne$ and $\ConvSatTwo$.

We acquire the estimates by the help of reference tones, that are transmitted and also received at the gateway station. From an operational point of view, these reference tones could be placed at transponder edges to preserve the spectral efficiency. Due to the different converter frequencies, the received tones could be clearly assigned to the individual satellites. Since all components in our gateway station have been synchronized by a reference clock, the frequency offset between the transmitted and the received pilot tones are solely caused by the link via the satellite's payload. The estimation must provide a sub-Hertz accuracy. To highlight this requirement, assume an estimation error of 1\unit{\Hz} for one diagonal entry of $\TCFO^{-1}$, while the other is perfectly estimated. The duration from the moment the reference tone arrives at the payload, back to the gateway station where $\TCFO^{-1}$ is calculated until the precompensated signal arrives again at the payload is typically around 250\unit{\ms}. The error of 1\unit{\Hz} would lead to a mismatched phase of 90\unit{\degree} which prohibits an efficient precoding, since phase uncertainties of a few degree already limit the \gls{ZF} precoding performance.
\newcommand{\offsetSatO}{\hat{f}_\text{Off1}}
\newcommand{\DopplerOneEst}{\widehat{\dopplerFreq_{\text{Sat1}}}}
\newcommand{\DopplerTwoEst}{\widehat{\dopplerFreq_{\text{Sat2}}}}
\newcommand{\ConvSatOneEst}{\widehat{f_{\text{Con1}}}}
\newcommand{\ConvSatTwoEst}{\widehat{f_{\text{Con2}}}}
\newcommand{\RefToneFreqO}{f_\text{Ref1}}

To evaluate the efficiency of our oscillator offset compensation, we calculate the remaining phase uncertainty after CFO precompensation with $\TCFO^{-1}$. Let $y_\text{ref1}(t)$ be the altered reference tone transmitted via satellite 1 and captured at the gateway station, modeled as
\begin{align}
	y_\text{Ref1}(t) = a(t) &\Euler^{-\complexLetter  \left( 2\pi (\RefToneFreqO + \DopplerOne + \ConvSatOne)  \cdot t  +  \pnSatO(t) + \phi_\text{Gw} (t) \right)} \nonumber \\
								&+ \noiseSymbol_\text{GW}(t)\
\end{align}
where $\RefToneFreqO$ is the frequency of the reference tone 1, the variable $\noiseSymbol_\text{GW}(t)$ denotes the noise and $a(t)$ is a slightly varying amplitude due to imperfections of the involved hardware. To acquire a nearly noiseless and normalized version of $y_\text{Ref1}(t)$, an \gls{ADPLL} with a low loop bandwidth of about 7\,Hz is locked to the signal. At the output of the ADPLL, we obtain
\begin{align}
	y_\text{PLL1}(t) \approx \Euler^{- \complexLetter  2\pi (\RefToneFreqO + \DopplerOneEst(t) + \ConvSatOneEst(t)) \cdot t }
\end{align}
with $\DopplerOneEst(t)$ and $\ConvSatOneEst(t)$ as estimates of the actual frequency offsets. Because of the low loop bandwith of the ADPLL, the phase noise entries $\widetilde{\pnSatO}(t)$ and $\widetilde{\phi_\text{Gw}}(t)$ are filtered out.\footnote{Potentially remaining phase noise at low offset frequencies could again be interpreted as the time-variant frequencies parts $\ConvSatOneEst(t)$ or $\ConvSatTwoEst(t)$, respectively.} Through the removal of the frequency part $\RefToneFreqO$ by a complex conjugate multiplication (see Fig.\ \ref{fig:sdr_impl}), the first element of $\TCFO^{-1}$ can be written as
\begin{align}
	\MatrixElement{\TCFO^{-1}}{1}{1} &= \Euler^{-\complexLetter 2\pi  \RefToneFreqO \cdot t} \cdot \overline{ y_\text{PLL1}(t)} \nonumber \\
	&= \Euler^{\complexLetter 2\pi \left( \DopplerOneEst(t) + \ConvSatOneEst(t) \right)  \cdot t},
\end{align}
where $\overline{ y_\text{PLL1}(t)}$ denotes the conjugate complex of $y_\text{PLL1}(t)$. To assess the efficiency of $\TCFO$, we compare $y_\text{PLL1}(t-250\unit{\ms})$ and $y_\text{Ref1}(t)$. The reasoning behind this approach is following: For a signal captured at time instant $t$ at the gateway station, a compensation of the oscillator offsets could have been applied based on the information available at time $t-250\unit{\ms}$. Hence, by comparing $y_\text{PLL1}(t-250\unit{\ms})$ and $y_\text{Ref1}(t)$, we provide an answer to the question: How large would the phase deviation have been, if $y_\text{Ref1}(t)$ had been precompensated based on the information of $y_\text{PLL1}(t-250\unit{\ms})$. The comparison between both signals can be written as
\begin{align}
	 \Delta_\text{Ref1}(t,\tau) &= y_\text{PLL1}(t-\tau) \cdot \overline{y_\text{Ref1}(t)} 																							\nonumber\\
        &= \Euler^{-\complexLetter 2\pi ( \DopplerOneEst(t-\tau) -\DopplerOne  + \ConvSatOneEst(t-\tau) - \ConvSatOne) \cdot t } \cdot				\nonumber \\
					&\hspace{1cm}		  \Euler^{-\complexLetter (\pnSatO(t) +  \phi_\text{Gw}(t)) } 					\nonumber \\
					 &= \Euler^{-\complexLetter ( 2\pi  \Delta f_\text{Sat1}(t,\tau)  \cdot t + \pnSatO(t) + \phi_\text{Gw}(t))}
\end{align}
Finally, the distortion introduced by up- and downconverter in the gateway is canceled out. The remaining frequency and phase uncertainty $\Delta{\phi_{\text{sys}}}$ after \gls{CFO} precompensation at the gateway is determined by
\begin{align}
&\Delta\phi_{\text{sys}}(t,\tau) 	= \arg\left( \Delta_\text{Ref1}(t,\tau) \cdot \overline{\Delta_\text{Ref2}(t,\tau)} \right)  \label{eqn:phaseUncertainty}\\
&									= \arg \left( \Euler^{- \complexLetter( 2\pi ( \Delta f_\text{Sat1}(t,\tau) - \Delta f_\text{Sat2}(t,\tau) ) \cdot t + \pnSatO(t) - \phi_\text{Sat2}(t))} \right)\nonumber
\end{align}
The components of $\Delta\phi_{\text{sys}}(t,\tau)$ are based on uncorrelated random processes because the satellites are separate entities. For this reason, a summation or subtraction of these components in (\ref{eqn:phaseUncertainty}) only increases the standard deviation of $\Delta\phi_{\text{sys}}(t,\tau)$.
\section{Results}
\label{sec:Results}
\subsection{CFO Compensation}

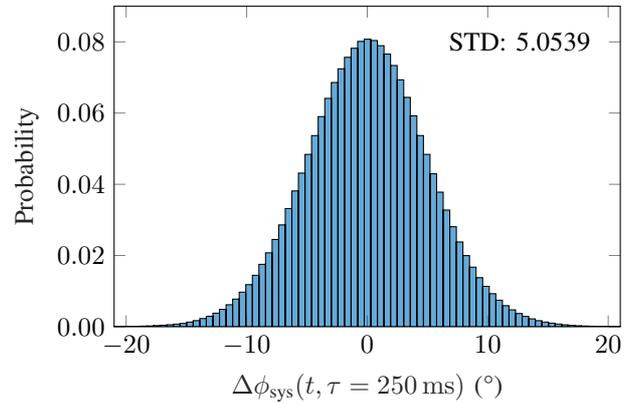
\begin{figure}[tb]
\centering
\input{plots/phase_diff_ref_tones}
\caption{Evaluation of the remaining phase uncertainty $\Delta\phi_{\text{sys}}(t,\tau)$ according to (\ref{eqn:phaseUncertainty}) after precompensation. Measurement over a time frame of two minutes.}
\label{fig:phase_uncertainty}
\end{figure}

In Fig.\ \ref{fig:phase_uncertainty} an unbiased histogram of $\Delta\phi_{\text{sys}}(t,\tau=250\unit{\ms})$ for a measurement over two minutes is plotted. Within this time frame, more than 37 million samples of $\Delta\phi_{\text{sys}}(t,\tau)$ have been collected and evaluated. It turns out that the remaining phase variations after the proposed precompensation follow a Gaussian distribution with a standard deviation of only 5\unit{\degree}. With this level of synchronization accuracy between the two satellites, \gls{MU-MIMO} precoding has become possible.
\subsection{Data transmission}
\begin{figure}[t]
	\centering
	\input{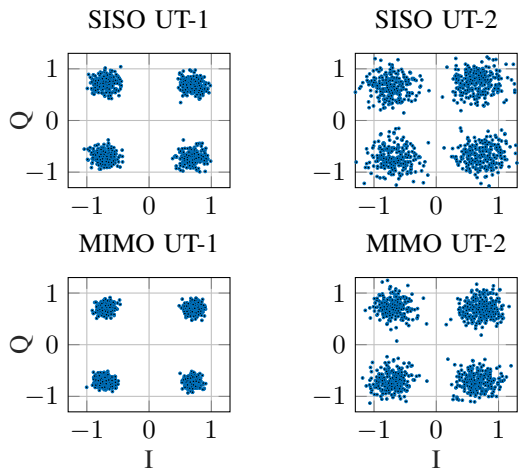}
	\caption{Comparison of decoded symbols as scatter plots for the during SISO and MIMO transmission mode.}
	\label{fig:scatter_comparison}
\end{figure}

To obtain a starting point for a comparison between \gls{SISO} and \gls{MU-MIMO} transmission, the trial started in SISO mode by setting $\precodM=\begin{bsmallmatrix}
1 & 0 \\
0 & 0 \\
\end{bsmallmatrix}$. In this state, both DVB-S2x receivers streamed out the same video content, namely Video Stream 1. Today, this is the expected use case since both \glspl{UT} receive the signal emitted by satellite 1 (broadcasting) whereas no signal is sent via satellite 2. Typically, satellite 2 acts as a spare satellite or is operated at a different "color", i.e.\ frequency or polarization.
We measured the \gls{MER} with spectrum analyzers and a known data sequence at both \glspl{UT}. In case of \gls{SISO} transmission, the \glspl{MER} measured are 16.5\unit{\dB} and 10.9\unit{\dB} for UT-1 and UT-2, respectively. The difference is caused by different antenna diameters i.e.\ receive \glspl{CNR}. A snapshot of 2000 decoded complex symbols during the SISO transmission is plotted in the first row of Fig.\ \ref{fig:scatter_comparison}.
In the second step, the precoding matrix $\precodM$ is calculated based on the \gls{CSI} provided \glspl{UT}. When $\precodM$ becomes a dense matrix, both satellites are utilized. Shortly after activating the \gls{MU-MIMO} mode, the DVB-S2x receiver at UT-2 begins a resynchronization and after a few seconds, the output has changed from Video Stream 1 to Video Stream 2. During that time, the DVB-S2x receiver at UT-1 continuously plays Video Stream 1.
To determine the signal quality with precoding activated, the MER has been measured again. For UT-1 the MER has become 18.3\unit{\dB}, and for UT-2 an MER of 12.3\unit{\dB} has been determined. A snapshot of 2000 decoded complex symbols captured during \gls{MU-MIMO} transmission is plotted in the second row of Fig.\ \ref{fig:scatter_comparison}.
\gls{MU-MIMO} precoding enabled the transmission of two independent signal streams to two non-cooperating single antenna receivers (unicast). Furthermore, the signal quality in terms of MER was improved at the receivers by 1.8\unit{\dB} and 1.4\unit{\dB}, respectively. This enhancement is the result of the constructive superposition of both satellite signals, i.e.\ the \gls{MIMO} \gls{SNR} gain.
Finally, if the \gls{MER} measurements are interpreted as conservative estimates for the \gls{SNR}, the channel capacity or rate can also be assessed. In case of \gls{SISO}, the best channel capacity was achieved with UT-1 leading to $R^\text{SISO}=\log_2(1+\text{SNR})=5.5 \unit{\bit/\s/\Hz}$. Using (\ref{eqn:sumrate}), the sum rate in case of ZF precoding is now $\sumRateZF = 6.1 + 4.2 = 10.3\unit{\bit/\s/\Hz}$, which is an improvement factor of nearly 1.9. If the \glspl{UT} were equipped with similar antenna diameters, this factor would be even be improved. Assuming two \glspl{UT} with similar values for $R^\text{SISO}$, both elements in $\sumRateZF$ would than provide an increased rate compared to the reference value $R^\text{SISO}$.

When comparing SISO and \gls{MU-MIMO}, it is important to point out that \gls{CFO} tracking and CSI estimation required additional resources. The produced overhead for \gls{CFO} tracking consisted of two continuous waves. \gls{CSI} was determined, on average, every second at the \glspl{UT}, resulting in an overhead of 2\,kBaud. The extra resources needed to enable \gls{MU-MIMO} are negligibly small in the light of the achievable gains and the fact, that a \gls{CRS} can be used simultaneously by all \glspl{UT}.

\section{Conclusion}
The paper presents the results of a field trial realizing a \gls{MU-MIMO} transmission via multiple satellites. To this end, the impact of \glspl{CFO} and oscillator phase noise have been accurately modeled. By the help of a reference system, the frequency offsets due to satellite motion and free-running oscillators was compensated remotely from the gateway station. Therewith, \gls{ZF} precoding was made possible and two independent data streams were transmitted to two single-antenna \glspl{UT} in the same frequency band and polarization (\gls{FFR} scheme). To demonstrate the compatibility with existing receiver hardware as well as the feasibility of \gls{MU-MIMO} precoding, we used \gls{COTS} DVB-S2x demodulators at the \glspl{UT} to process the incoming signals. \gls{MER} improvements of about 1.5\, dB and a rate enhancement of 90\,\% demonstrate the success of this first precoding test campaign for communication satellites. The results prove that \gls{MIMO} precoding is possible with existing satellites in orbit, even in the more complex multiple-reflector setup that involves the signal phase. The results can the be transferred to multi-group or multi-cast precoding in a straightforward manner.

\bibliographystyle{IEEEtran}
\bibliography{ref,ref2}

\end{document}

%% file: imports.tex
\usepackage{algorithmic}

\usepackage[english]{babel}
\usepackage{blindtext}	

\usepackage{nicefrac}

\usepackage{amsmath,amssymb,amsfonts}
\usepackage{amsthm}           
\usepackage{bm}               
\usepackage{stmaryrd}         
\usepackage{wasysym}          
\usepackage{boxedminipage}    
\usepackage{fancyhdr}         
\usepackage{lastpage}         
\usepackage[dvips]{epsfig}    
\usepackage{array}            
\usepackage{multirow}         
\usepackage{epic}             
\usepackage{color}            
\usepackage{listings}         
\usepackage{float}            
\usepackage{rotfloat}            
\usepackage{longtable}        
\usepackage{wrapfig}          
\usepackage{colortbl}
\usepackage{graphicx}
\usepackage{tabularx}

\usepackage{mathtools}

\usepackage[detect-weight=true, binary-units=true]{siunitx}

\usepackage{pst-all}

\usepackage{textcomp} 
\usepackage{glossaries}

\usepackage{framed}
\usepackage{xcolor}
\colorlet{shadecolor}{gray!15}

\usepackage[noadjust]{cite}

\usepackage{tikz}
\usepackage{pgfplots}
\pgfplotsset{compat=newest}
\usetikzlibrary{shapes,arrows, arrows.meta}
\usetikzlibrary{positioning}
\usetikzlibrary{calc}
\usetikzlibrary{fit}
\usetikzlibrary{patterns,decorations.pathreplacing}
\usetikzlibrary{spy,backgrounds}

\usepackage{rotating}



%% file: commands.tex
\usepackage{amsmath}

\newcommand{\Vector}[1]{\boldsymbol{\mathbf{#1}}}    
\newcommand{\Matrix}[1]{\boldsymbol{\mathbf{#1}}}

\newcommand{\recSymbol}{y}
\newcommand{\recVec}{\Vector{\recSymbol}}

\newcommand{\dataSymbol}{d}
\newcommand{\dataVec}{\Vector{\dataSymbol}}

\newcommand{\sendSymbol}{x}
\newcommand{\sendVec}{\Vector{\sendSymbol}}

\newcommand{\noiseSymbol}{n}
\newcommand{\noiseVec}{\Vector{\noiseSymbol}}

\newcommand{\channelSymbol}{h}
\newcommand{\ctm}{\Matrix{\MakeUppercase{\channelSymbol}}}
\newcommand{\channelVec}{\Vector{\channelSymbol}}

\newcommand{\ctmt}{\tilde{\ctm}}

\newcommand{\precodSymbol}{p}
\newcommand{\precodVec}{\Vector{\precodSymbol}}
\newcommand{\precodM}{\Matrix{\MakeUppercase{\precodSymbol}}}

\newcommand{\Ntran}{N}

\newcommand{\Nusers}{K}

\newcommand{\antennaDist}[2]{r_{#1,#2}}

\newcommand{\complexSet}[1]{\mathbb{C}^{#1}}

\newcommand{\complexLetter}{j}

\newcommand{\speedOfLight}{c_0}
\newcommand{\Euler}{\mathrm{e}}

\newcommand{\Hermitian}[1]{{#1}^{H}}
\newcommand{\Transpose}[1]{{#1}^{T}}

\newcommand{\unit}[1]{\,\si{#1}}
\newcommand{\diag}[1]{\text{diag}\left( #1 \right)}

\newcommand{\MatrixElement}[3]{  {\left[ {#1} \right]}_{#2,#3} }

\makeatletter
\newcommand*{\rom}[1]{\expandafter\@slowromancap\romannumeral #1@}
\makeatother

\newcount\colveccount
\newcommand*\colvec[1]{
        \global\colveccount#1
        \begin{bmatrix}
        \colvecnext
}
\def\colvecnext#1{
        #1
        \global\advance\colveccount-1
        \ifnum\colveccount>0
                \\
                \expandafter\colvecnext
        \else
                \end{bmatrix}
        \fi
}

\newcommand{\LnbFreqUserOne}{f_{\text{LNB1}}(t)}
\newcommand{\LnbFreqUserTwo}{f_{\text{LNB2}}(t)}
\newcommand{\dopplerFreq}{\nu}
\newcommand{\DopplerOne}{\dopplerFreq_{\text{Sat1}}\left(t\right)}
\newcommand{\DopplerTwo}{\dopplerFreq_{\text{Sat2}}\left(t\right)}

\newcommand{\ConvSatOne}{f_{\text{Con1}}(t)}
\newcommand{\ConvSatTwo}{f_{\text{Con2}}(t)}

\newcommand{\pnSatO}{\phi_\text{Sat1}}
\newcommand{\pnSatT}{\phi_\text{Sat2}}
\newcommand{\UplinkFreqOne}{f^{\text{U}}_{\text{Sat1}}}
\newcommand{\UplinkFreqTwo}{f^{\text{U}}_{\text{Sat2}}}
\newcommand{\DownlinkFreqOne}{f^{\text{D}}}
\newcommand{\TCFO}{\Matrix{T}(t)}
\newcommand{\RCFO}{\Matrix{R}(t)}

%% file: gloss.tex
\newacronym[longplural={High Throughput Satellites}]{HTS}{HTS}{High Throughput Satellite}
\newacronym{ADPLL}{ADPLL}{all digital phase locked loop}
\newacronym{APC}{PAPC}{per-antenna power constraint}
\newacronym{AP}{AP}{Access Point}
\newacronym{CCDF}{CCDF}{complementary cumulative distribution function}
\newacronym{CCI}{CCI}{co-channel interference}
\newacronym{CFO}{CFO}{carrier frequency offset}
\newacronym{CI}{CI}{channel inversion}
\newacronym{CoC}{CoC}{coefficient of correlation}
\newacronym{COTS}{COTS}{commercial off-the-shelf}
\newacronym{CSI}{CSI}{channel state information}
\newacronym{CTM}{CTM}{channel transfer matrix}
\newacronym{CNR}{CNR}{carrier-to-noise ratio}
\newacronym{DPC}{DPC}{dirty paper coding}
\newacronym{EHF}{EHF}{extremely high frequency}
\newacronym{FDMA}{FDMA}{frequency division multiple access}
\newacronym{FFR}{FFR}{full frequency reuse}
\newacronym{FL}{FL}{forward link}
\newacronym{FR4}{FR4}{four-color frequency re-use}
\newacronym{FSS}{FSS}{fixed satellite services}
\newacronym{GEO}{GEO}{geosynchronous equatorial orbit}
\newacronym{LHCP}{LHCP}{left hand circular polarisation}
\newacronym{LNB}{LNB}{low-noise block downconverter}
\newacronym{LOS}{LOS}{line-of-sight}
\newacronym{MADOC}{MADOC}{multiple antenna downlink orthogonal clustering}
\newacronym{MER}{MER}{modulation error ratio}
\newacronym{MIMO}{MIMO}{multiple-input multiple-output}
\newacronym{MISO}{MISO}{multiple-input single-output}
\newacronym{MMSE}{MMSE}{minimum mean square error}
\newacronym{MPI}{MPI}{Moore-Penrose-Inverse}
\newacronym{MU-MIMO}{MU-MIMO}{multi-user \acrlong{MIMO}}
\newacronym{RCI}{RCI}{regularized channel inversion}
\newacronym{RF}{RF}{radio frequency}
\newacronym{RHCP}{RHCP}{right hand circular polarisation}
\newacronym{RMS}{rms}{root mean square}
\newacronym{SATCOM}{SATCOM}{satellite communications}
\newacronym{SDMA}{SDMA}{space division multiple access}
\newacronym{SDR}{SDR}{software-defined radio}
\newacronym{SFPB}{SFPB}{single-feed-per-beam}
\newacronym{SINR}{SINR}{signal-to-interference-and-noise ratio}
\newacronym{SISO}{SISO}{single-input single-output}
\newacronym{SNR}{SNR}{signal-to-noise ratio}
\newacronym{SPC}{SPC}{sum power constraint}
\newacronym{SVD}{SVD}{singular value Decomposition}
\newacronym{TDMA}{TDMA}{time division multiple access}
\newacronym{UCA}{UCA}{uniform circular array}
\newacronym{UE}{UE}{user equipment}
\newacronym{UT}{UT}{user terminal}
\newacronym{VHTS}{VHTS}{Very High Throughput Satellite}
\newacronym{VOD}{VOD}{video-on-demand}
\newacronym{VSAT}{VSAT}{very small aperture terminal}
\newacronym{ZF}{ZF}{zero forcing}
\newacronym{RTT}{RTT}{round trip time}
\newacronym{CRS}{CRS}{common reference signal}

%% file: pics/scenario.tex

\tikzset{every picture/.style={thick}}

\begin{tikzpicture}
	
  \node [anchor=south west] (label) at (0,0) {\includegraphics[width = \columnwidth,trim=20 0 100 0, clip]{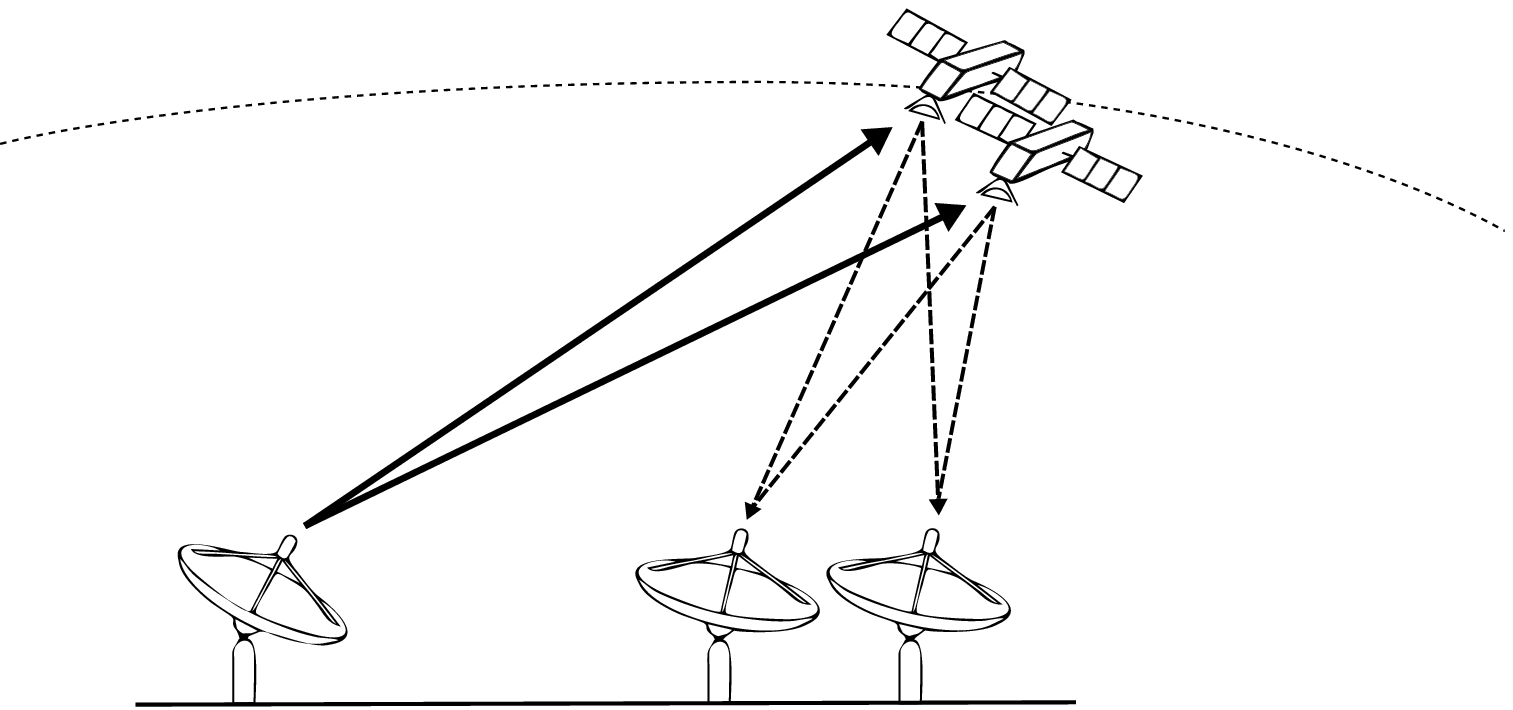}};

  	\node [rectangle] at (1.5,-0.05) {Gateway};
	\node [rectangle] at (5.5,-0.05) {UT-1};
	\node [rectangle] at (7.0,-0.05) {UT-2};

	\node [rectangle] at (6.4,5.5) {E7A};
	\node [rectangle] at (8.4,3.9) {E7B};

	\node  [rectangle] at (6.8,2.6) { $\ctm$};

	\node [rotate = 4] at (3,5.3) {\scriptsize Geostationary Orbit};


	

	
	
	
	
	
\end{tikzpicture}%

%% file: pics/antennafarm.tex

\tikzset{every picture/.style={thick}}

\begin{tikzpicture}[every node/.style={inner sep=1,outer sep=1}, font=\scriptsize]

	\node [anchor=south west] (label) at (0,0) {\includegraphics[width = \columnwidth,trim=0 0 0 0, clip]{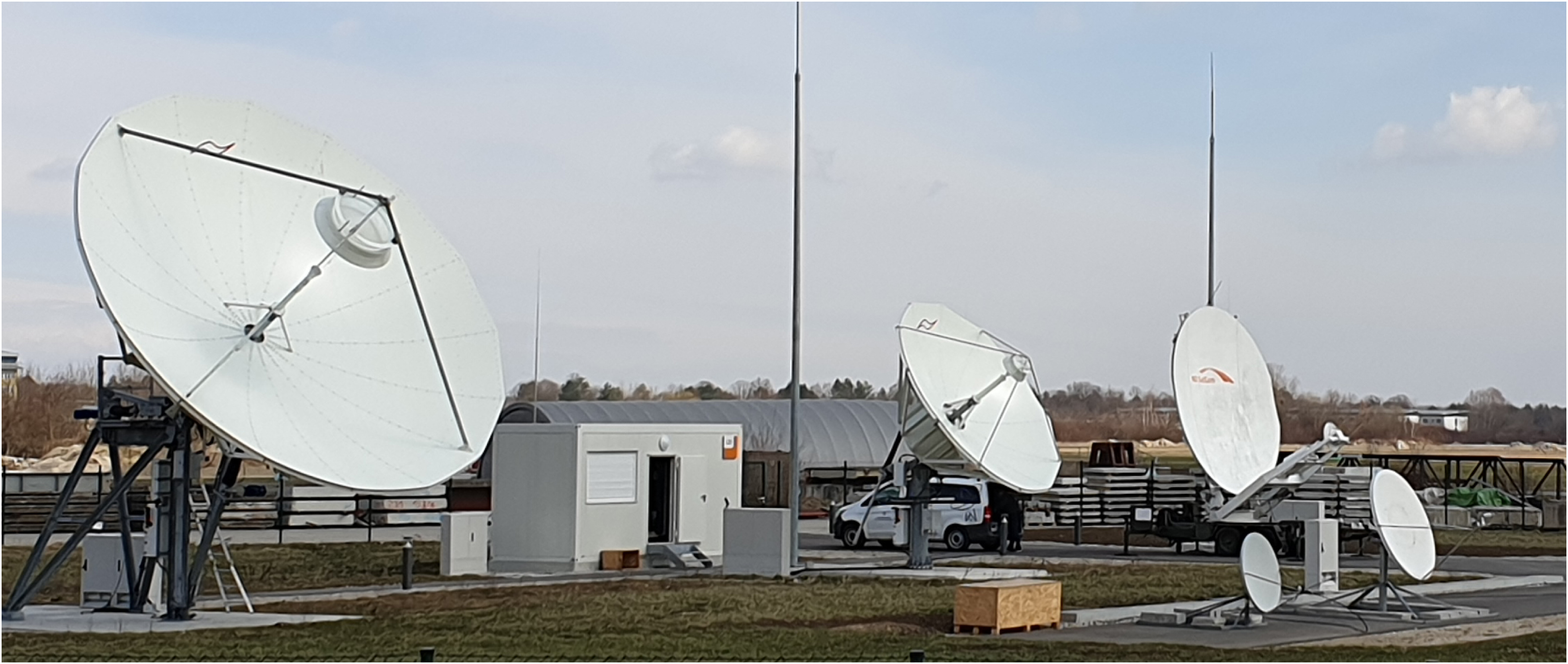}};
  \node [fill=white] at (1.4,3.5) {UT-1  (\O\,7.6m)};
  \node [rectangle, fill=white] at (6.8,2.4) {Gateway  (\O\,4.6m)};
  \node [rectangle, fill=white] at (8.15,0.3) {UT-2  (\O\,1.8m)};

\end{tikzpicture}

%% file: pics/sdr_impl.tex

%

	\tikzstyle{line}    = [draw, line width=.5pt]
	\tikzstyle{lineA}    = [line, -latex]
	\tikzstyle{lineR}    = [line, latex-]
	\tikzstyle{block}     = [draw,  rectangle, minimum height=3.5mm, minimum width=1.6em]

	\tikzstyle{sum}       = [draw, circle, line]
	\tikzstyle{mux}     = [draw, rectangle, minimum height=.8cm, minimum width=1.5mm]

	\tikzstyle{condot} 	= [fill=black, shape=circle, inner sep=1pt, radius=.6mm]

	\newcommand{\drawMult}[1] {
	  	\draw [line,-]	(#1)      +(-1.25mm,-1.25mm) -- +(1.25mm,1.25mm);
	    \draw [line,-]	(#1)      +(-1.25mm,1.25mm) -- +(1.25mm,-1.25mm);
	}

	\newcommand{\drawSum}[1] {
	  	\draw [line,-]	(#1)      +(-1.25mm,0) -- +(1.25mm,0);
	    \draw [line,-]	(#1)      +(0,1.25mm) -- +(0,-1.25mm);
	}

   \centering
    \begin{tikzpicture}[font=\scriptsize, node distance = 0.7cm]

    	\node [block, align=center]												               (Stream1)		{Video\\Stream 1};
    	\node [block, above=0.4 of Stream1, align=center]				                       (DVBO)			{DVB-S2x\\Mod};
    	\node [block, right= .6cm of Stream1, align=center]						               (Stream2)		{Video\\Stream 2};
    	\node [block, above= 0.4 of Stream2, align=center]				                       (DVBT)			{DVB-S2x\\Mod};

    	\node (helper) at ($(DVBO)!0.5!(DVBT)$) {};

    	\node [block, above=0.6 of helper, minimum width=3cm, minimum height = .6cm] 			   (precoder) 		{$\precodM$};
        \node [block, above=0.5 of precoder, align=center]                                     (mimopilot)      {MIMO\\Pilots};

        \node [sum] at (mimopilot-|Stream1) (pilotSumO) {};
        \node [sum] at (mimopilot-|Stream2) (pilotSumT) {};

        \drawSum{pilotSumO};
        \drawSum{pilotSumT};

        \draw[lineA] (mimopilot) -- (pilotSumO);
        \draw[lineA] (mimopilot) -- (pilotSumT);

    	\node [block, above=2.5cm of DVBO]							(CFOCombO)		{$\MatrixElement{\TCFO^{-1}}{1}{1}$};
    	\node [block, above=3.1 of DVBT]							(CFOCombT)		{$\MatrixElement{\TCFO^{-1}}{2}{2}$};

    	\node [block, above right=-.6cm and 0.6cm of CFOCombT, align=center]		(RefToneGen)	{Ref.\\Tone\\Gen};
    	\node [condot, right = 1mm of RefToneGen.east]								(RefToneEast) {};
    	\node [condot, left = 1mm of RefToneGen.west]								(RefToneWest) {};

    	\node [sum, above=0.8 of CFOCombO]						(toneAddO)	{};
    	\node [sum, above=0.7 of CFOCombT]						(toneAddT)	{};

    	\drawSum{toneAddO};
    	\drawSum{toneAddT};

    	\draw [lineA] (Stream1) -- (DVBO);
    	\draw [lineA] (Stream2) -- (DVBT);
    	\draw [lineA] (DVBO) -- (DVBO.north|-precoder.south);
    	\draw [lineA] (DVBT) -- (DVBT.north|-precoder.south);

		\draw [lineA] (DVBO.north|-precoder.north) -- (pilotSumO);
		\draw [lineA] (DVBT.north|-precoder.north) -- (pilotSumT);

        \draw [lineA] (pilotSumO) -- (CFOCombO);
        \draw [lineA] (pilotSumT) -- (CFOCombT);

		\draw [line] (RefToneGen.east) -- (RefToneEast);
		\draw [line] (RefToneGen.west) -- (RefToneWest);
    	
    	\draw [lineA] (RefToneEast) |- (toneAddT);
    	
    	\draw [lineA] (RefToneWest) |- (toneAddO);

    	\node [sum, above right= 0.9 and 0.5cm of toneAddO]								(mixerO)	{};
    	\drawMult{mixerO};
    	\node [left=0.4cm of mixerO]								(mixerOinput)	{$\Euler^{-\complexLetter 2\pi\UplinkFreqOne t}$};
    	
    	\draw [lineA] (mixerOinput) -- (mixerO);
    	
    	\node [sum ]												(mixerT) 		at (toneAddT|-mixerO)	{};
    	\drawMult{mixerT};
    	\node [right=0.4cm of mixerT]								(mixerTinput)	{$\Euler^{-\complexLetter 2\pi\UplinkFreqTwo t}$};
    	\draw [lineA] (mixerTinput) -- (mixerT);

    	\node [sum, above=0.3cm of mixerT] 								(txSum)		{};
    	\drawSum{txSum};

    	\node [block, above right= .3cm and 1cm of txSum, align=center]								(dish)		{Gateway Antenna};

		\draw [lineA] (CFOCombO) -- (toneAddO);
		\draw [lineA] (CFOCombT) -- (toneAddT);

		\draw [lineA] (toneAddO) -- ++(0,.5) -| (mixerO);
		\draw [lineA] (toneAddT) -- (mixerT);

    	\draw [lineA] (mixerT) -- (txSum);
    	\draw [lineA] (mixerO) |- (txSum);
    	\draw [lineA] (txSum) |- (dish);


    	\node [block, right= 0.4cm of RefToneGen]							(adpllO) 	{ADPLL};
    	\node [sum, below=0.4cm of adpllO]								(conjO)		{};
    	\drawMult{conjO};

    	\draw [lineA] (adpllO) -- node[left=-2pt, pos=0.75] {*} (conjO);

    	\node [block, right= 0.3cm of adpllO]						(adpllT) 	{ADPLL};
    	\node [sum, below=of adpllT]								(conjT)		{};
		\drawMult{conjT};

		\draw [lineA] (adpllT) -- node[left=-2pt, pos=0.8] {*} (conjT);

		\draw [lineA] (RefToneEast) |- (conjO);
		\draw [lineA] (RefToneWest) |- (conjT);


        \draw [lineA, dashed] (conjO.south) -- ++(0,-.6) -- ++(-2cm,0) |- (CFOCombT.east);
		\draw [lineA, dashed] (conjT.south) -- ++(0,-.5) -- ++(-4cm,0) |- (CFOCombO.east);

		\node (helper2) at ($(adpllO)!0.5!(adpllT)$) {};
		
		\node [sum] (rxmixer) at (helper2|-txSum) {};
    	\node [left=0.4cm of rxmixer]		(mixerRxinput)	{$\Euler^{-\complexLetter 2\pi\DownlinkFreqOne t}$};
    	\drawMult{rxmixer};

    	\draw [lineA] (mixerRxinput) -- (rxmixer);

		\draw [lineA] (dish.east) -| (rxmixer);

		\node [condot, below=1.4 of rxmixer] 						(condotrx)  {};

		\draw [lineA] (rxmixer) -- (condotrx);
		\draw [lineA] (condotrx) -| (adpllO);
		\draw [lineA] (condotrx) -| (adpllT);

		\draw [lineR, dashed] (precoder.6) -- node[pos=0.5, fill=blue!2] {CSI UT-1} ++(5cm,0);
		\draw [lineR, dashed] (precoder.354) -- node[pos=0.5, fill=blue!2] {CSI UT-2} ++(5cm,0);

    \begin{pgfonlayer}{background}

    	   	
	   	   	\node[draw, fill=blue!2,rounded corners, draw=black!50, dashed, 
     			fit={([xshift=-1mm,yshift=0mm]CFOCombO.west|-toneAddT.north) ([yshift=-1mm]Stream1.south-|adpllT.east)},
     				label={[anchor=south,inner sep=1pt]south:\hspace{3cm} Transmit SDR}] () {};

    \end{pgfonlayer}

	\end{tikzpicture}


%% file: osci_chain.tex




	\tikzstyle{line}    = [draw, line width=.5pt]
	\tikzstyle{lineA}    = [line, -latex]
	\tikzstyle{lineR}    = [line, latex-]
	\tikzstyle{block}     = [draw,  rectangle, minimum height=3.5mm, minimum width=1.6em]
	\tikzstyle{block}     = [draw,  rectangle, minimum height=3.5mm, minimum width=1.6em]

	\tikzstyle{sum}       = [draw, circle, line]
	\tikzstyle{mux}     = [draw, rectangle, minimum height=.8cm, minimum width=1.5mm]

	\tikzstyle{condot} 	= [fill=black, shape=circle, inner sep=1pt, radius=.6mm]

	\newcommand{\drawMult}[1] {
	  	\draw [line,-]	(#1)      +(-1.25mm,-1.25mm) -- +(1.25mm,1.25mm);
	    \draw [line,-]	(#1)      +(-1.25mm,1.25mm) -- +(1.25mm,-1.25mm);
	}

	\newcommand{\drawSum}[1] {
	  	\draw [line,-]	(#1)      +(-1.25mm,0) -- +(1.25mm,0);
	    \draw [line,-]	(#1)      +(0,1.25mm) -- +(0,-1.25mm);
	}

   \centering
    \begin{tikzpicture}[font=\scriptsize, node distance = 0.7cm]

    	\node [block, minimum height=40.5mm]	(sdr) 	{SDR};
    	
    	\node [sum, right= of sdr.72]			(mix1)	{};
    	\drawMult{mix1};
		\node [above= 0.4cm of mix1] 			(mix1Input) {$\Euler^{ -\complexLetter 2 \pi \UplinkFreqOne\cdot t}$};
		\draw [lineA] (mix1Input) -- (mix1);

    	\node [sum, right= of sdr.42]			(mix2)	{};
    	\drawMult{mix2};
		\node [below= 0.4cm of mix2] 			(mix2Input) {$\Euler^{ -\complexLetter 2 \pi \UplinkFreqTwo\cdot t}$};
		\draw [lineA] (mix2Input) -- (mix2);

    	\node [sum, right= 0.5cm of  $(mix1)!0.5!(mix2)$] (sumTx) {};
    	\drawSum{sumTx};

    	\node [condot, right = 0.4cm of sumTx]		(condot1)	{};
    	\node [block, above right=.35cm and .8cm of condot1]	(H1Uplink)	{$\channelSymbol^{\text{U}}_1$};
    	\node [block, below right=.35cm and .8cm of condot1]	(H2Uplink)	{$\channelSymbol^{\text{U}}_2$};

		\draw [lineA] (sdr.east|-mix1) -- (mix1);
    	\draw [lineA] (sdr.east|-mix2) -- (mix2);

    	\draw [lineA] (mix1) -| (sumTx);
    	\draw [lineA] (mix2) -| (sumTx);

    	\draw [line] (sumTx) -- (condot1);


    	\node [sum, right= 2.1cm of H1Uplink]		(dopplerMultU1) {};
    	\drawMult{dopplerMultU1};

		\node [above= 0.4cm of dopplerMultU1] 	(u1DopplerInput) {$\Euler^{ -\complexLetter \left( 2 \pi \left(\DopplerOne + \ConvSatOne\right)\cdot t + \pnSatO(t) \right) }$};




    	\node [sum, right= 2.1cm of H2Uplink]		(dopplerMultU2) {};
    	\drawMult{dopplerMultU2};
    	\node [below= 0.4cm of dopplerMultU2] 	(u2DopplerInput) {$\Euler^{ -\complexLetter \left( 2 \pi \left(\DopplerTwo + \ConvSatTwo\right)\cdot t + \pnSatT(t) \right)}$};



    	\node [block, right= 7.0cm of condot1, minimum height=20.5mm]	(Hd) 	{$\ctm$};

    	\node [condot, left = 0.6cm of dopplerMultU1-|Hd.west]		(condot2a)	{};
    	\node [condot, left = 0.25cm of dopplerMultU2-|Hd.west]		(condot2b)	{};

    	\node [block, below left= 2.6cm and 1cm of condot2a] (H1Downlink) {$\channelSymbol^{\text{D}}_1$};
    	\node [block, below = 0.4cm of H1Downlink] (H2Downlink) {$\channelSymbol^{\text{D}}_2$};

    	\node [sum, left = of $(H1Downlink)!0.5!(H2Downlink)$ ] (obsSum) {};
    	\drawSum{obsSum};
		
		\node [sum] at ([xshift=1mm]sumTx|-obsSum) (LNB) {};
		\drawMult{LNB};
		\node [above= .4cm of LNB] 			(lnbInput) {$\Euler^{ -\complexLetter \left( 2 \pi \DownlinkFreqOne \cdot t + \phi_\text{Gw}(t) \right)}$};

		\node [sum, right = 1.6 of dopplerMultU1-|Hd.east]		(LnbR1) {};
		\drawMult{LnbR1};
		\node [above= .4cm of LnbR1] 			(LnbR1Input) {$\Euler^{ -\complexLetter \left( 2 \pi \LnbFreqUserOne \cdot t + \phi_\text{LNB1}(t) \right)  }$};
		\draw [lineA] (LnbR1Input) -- (LnbR1);

		\node [condot, right= 0.3 of LnbR1] 						(connDotR1) {};

		\node [block, above right = .1mm and 0.5cm of connDotR1]		(DVBS2R1) {DVB-S2x Receiver};
		\node [block, below right = .1mm and 0.5cm of connDotR1]		(SDRR1) {SDR};

		\node [sum, right = 1.6 of dopplerMultU2-|Hd.east]		(LnbR2) {};
		\drawMult{LnbR2};
		\node [condot, right= 0.3 of LnbR2] 						(connDotR2) {};
		\node [below= .4cm of LnbR2] 			(LnbR2Input) {$\Euler^{ -\complexLetter \left( 2 \pi \LnbFreqUserTwo \cdot t + \phi_\text{LNB2}(t) \right) }$};
		\draw [lineA] (LnbR2Input) -- (LnbR2);

		\node [block, above right = .1mm and 0.5cm of connDotR2]		(DVBS2R2) {DVB-S2x Receiver};
		\node [block, below right = .1mm and 0.5cm of connDotR2]		(SDRR2) {SDR};


    	\draw[lineA] (condot1) |- (H1Uplink);
    	\draw[lineA] (condot1) |- (H2Uplink);

    	\draw[lineA] (H1Uplink) -- (dopplerMultU1);
    	\draw[lineA] (u1DopplerInput) -- (dopplerMultU1);


    	\draw[lineA] (H2Uplink) -- (dopplerMultU2);
    	\draw[lineA] (u2DopplerInput) -- (dopplerMultU2);



    	\draw[lineA] 	(dopplerMultU1.east) -- (dopplerMultU1-|Hd.west);
    	\draw[lineA] 	(dopplerMultU2.east) -- (dopplerMultU2-|Hd.west);

    	\draw [lineA] (condot2a) |- (H1Downlink);
    	\draw [lineA] (condot2b) |- (H2Downlink);

    	\draw [lineA] (H1Downlink) -| (obsSum);
    	\draw [lineA] (H2Downlink) -| (obsSum);

		\draw [lineA] (obsSum) -- (LNB);
		\draw [lineR] (sdr.290) -- ([xshift=3mm]sdr.290) |- (LNB);
		\draw [lineA] (lnbInput) -- (LNB);

		\draw [lineA] 	(dopplerMultU1-|Hd.east) -- (LnbR1);
		\draw [line] 	(LnbR1) -- (connDotR1);
		\draw [lineA] 	(connDotR1) |- (DVBS2R1);
		\draw [lineA] 	(connDotR1) |- (SDRR1);

		\draw [lineA] 	(dopplerMultU2-|Hd.east) -- (LnbR2);
		\draw [line] 	(LnbR2) -- (connDotR2);
		\draw [lineA] 	(connDotR2) |- (DVBS2R2);
		\draw [lineA] 	(connDotR2) |- (SDRR2);

    \begin{pgfonlayer}{background}

     	\node[draw, fill=blue!2,rounded corners, draw=black!50, dashed, 
     			fit={([xshift=-1mm]sdr.west|- mix1Input.north) ([xshift=-1.5mm, yshift=-2mm] lnbInput.east|-LNB.south)},
     				label={[anchor=south,inner sep=0pt]south:Gateway Station}] () {};

     	\node[draw, fill=blue!2,rounded corners, draw=black!50, dashed, 
     			fit={([xshift=1mm,yshift=-2mm]u1DopplerInput.north west) ([yshift=-2mm]u1DopplerInput.north east|-dopplerMultU1.south east)},
     				label={[anchor=south,inner sep=1pt]south:Satellite 1}] () {};

     	\node[draw, fill=blue!2,rounded corners, draw=black!50, dashed, 
     			fit={([xshift=1mm,yshift=2mm]u2DopplerInput.west|-dopplerMultU2.north) ([yshift=1mm]u2DopplerInput.south east)},
     				label={[anchor=north,inner sep=1pt]north:Satellite 2}] () {};

     	\node[draw, fill=blue!2,rounded corners, draw=black!50, dashed, 
     			fit={([xshift=1mm,yshift=-1mm] LnbR1Input.north west) ([xshift=4mm, yshift=0mm]DVBS2R1.east|-SDRR1.south)},
     				label={[anchor=north,inner sep=1pt]north:\hspace{3.0cm} UT-1}] () {};



     	\node[draw, fill=blue!2,rounded corners, draw=black!50, dashed, 
     			fit={([xshift=1mm,yshift=1mm]LnbR2Input.south west) ([xshift=4mm, yshift=0mm]DVBS2R2.north east)},
     				label={[anchor=south,inner sep=1pt]south:\hspace{3.0cm} UT-2}] () {};

    \end{pgfonlayer}

	\end{tikzpicture}


%% file: plots/phase_diff_ref_tones.tex
%
%
\definecolor{mycolor1}{rgb}{0.00000,0.44700,0.74100}%
\begin{tikzpicture}

\begin{axis}[%
width=0.951\figurewidth,
height=\figureheight,
at={(0\figurewidth,0\figureheight)},
scale only axis,
xmin=-21,
xmax=21,
xlabel style={font=\color{white!15!black}},
xlabel={$\Delta\phi_{\text{sys}}(t,\tau=250\unit{\ms})$ (\si{\degree})},
ymin=0,
ymax=0.09,
ylabel style={font=\color{white!15!black}},
ylabel={Probability},
axis background/.style={fill=white},
scaled y ticks = false,
y tick label style={/pgf/number format/.cd, fixed, fixed zerofill,precision=2}
]
\addplot[ybar interval, fill=mycolor1, fill opacity=0.6, draw=black, area legend] table[row sep=crcr] {%
x	y\\
-27	3.90720390720391e-07\\
-26.454	4.88400488400489e-07\\
-25.908	3.90720390720389e-07\\
-25.362	1.02564102564103e-06\\
-24.816	3.76068376068377e-06\\
-24.27	4.44444444444445e-06\\
-23.724	4.73748473748474e-06\\
-23.178	5.95848595848593e-06\\
-22.632	1.13308913308913e-05\\
-22.086	1.77777777777778e-05\\
-21.54	2.3931623931624e-05\\
-20.994	3.81929181929182e-05\\
-20.448	5.73382173382174e-05\\
-19.902	7.71672771672772e-05\\
-19.356	0.000114090354090353\\
-18.81	0.000172454212454213\\
-18.264	0.000218608058608059\\
-17.718	0.000315213675213673\\
-17.172	0.000423150183150184\\
-16.626	0.00055941391941392\\
-16.08	0.000731428571428572\\
-15.534	0.000945152625152626\\
-14.988	0.00129777777777777\\
-14.442	0.00171335775335776\\
-13.896	0.00229313797313798\\
-13.35	0.00293479853479853\\
-12.804	0.00379921855921856\\
-12.258	0.00488634920634921\\
-11.712	0.006170989010989\\
-11.166	0.0077523321123321\\
-10.62	0.00969968253968255\\
-10.074	0.0118240293040293\\
-9.528	0.0144375091575092\\
-8.982	0.0175188278388279\\
-8.436	0.0207482783882783\\
-7.89	0.0245018803418804\\
-7.344	0.0285924297924298\\
-6.798	0.033185054945055\\
-6.252	0.0381256654456655\\
-5.706	0.0431411965811963\\
-5.16	0.0483888156288157\\
-4.614	0.0536737484737485\\
-4.068	0.058999413919414\\
-3.522	0.0641306959706961\\
-2.976	0.0685863736263737\\
-2.43	0.0724213431013427\\
-1.884	0.0756589010989012\\
-1.338	0.0782068376068377\\
-0.791999999999998	0.0801307448107449\\
-0.245999999999999	0.0807473992673994\\
0.300000000000001	0.0803771916971913\\
0.846000000000004	0.0789263980463981\\
1.392	0.0765674236874238\\
1.938	0.0733602442002443\\
2.484	0.0693243956043957\\
3.03	0.0644127960927962\\
3.576	0.0592126495726492\\
4.122	0.0536905982905984\\
4.668	0.0483940415140413\\
5.21400000000001	0.0429513553113554\\
5.76000000000001	0.0377965811965812\\
6.306	0.0327604884004884\\
6.852	0.0280798046398047\\
7.398	0.0238007814407815\\
7.944	0.0199799267399268\\
8.49	0.0167224908424909\\
9.036	0.0137783638583639\\
9.582	0.011336507936508\\
10.128	0.009273894993895\\
10.674	0.00743921855921847\\
11.22	0.00590495726495727\\
11.766	0.0047376800976801\\
12.312	0.00371448107448108\\
12.858	0.00289738705738706\\
13.404	0.00220986568986569\\
13.95	0.0016979242979243\\
14.496	0.00130163614163614\\
15.042	0.00100874236874237\\
15.588	0.000750427350427351\\
16.134	0.000540122100122094\\
16.68	0.000384322344322345\\
17.226	0.000261440781440782\\
17.772	0.000195360195360196\\
18.318	0.000140415140415141\\
18.864	9.18192918192919e-05\\
19.41	6.24664224664225e-05\\
19.956	4.33699633699634e-05\\
20.502	3.37484737484738e-05\\
21.048	1.6996336996337e-05\\
21.594	1.04029304029304e-05\\
22.14	8.10744810744801e-06\\
22.686	5.56776556776557e-06\\
23.232	3.61416361416362e-06\\
23.778	1.26984126984127e-06\\
24.324	1.46520146520147e-07\\
24.87	1.46520146520147e-07\\
25.416	1.46520146520147e-07\\
25.962	1.95360195360196e-07\\
26.508	1.95360195360196e-07\\
27.054	4.88400488400489e-07\\
27.6	4.88400488400489e-07\\
};
\node[right, align=left]
at (axis cs:6,0.08) {STD: 5.0539};
\end{axis}
\end{tikzpicture}%